\documentclass[11pt,epsfig]{article}
\newcommand{\be}{\begin{equation}}
\newcommand{\ee}{\end{equation}}
\newcommand{\beqn}{\begin{eqnarray}}
\newcommand{\eeqn}{\end{eqnarray}}
\begin{document}
\noindent
DESY 99-139\\                           
hep-ph/9909344\\[1cm]
\begin{center}
\begin{LARGE}
{\bf The small-x Behavior of the Nonsinglet Polarized Structure 
Function g$_2$  
in the Double Logarithmic Approximation}\\[1cm]
\end{LARGE}
J.Bartels\footnote{Supported by the TMR Network ``QCD and Deep Structure of 
Elementary Particles''}\\
II.Institut f.Theoretische Physik Universit\"at Hamburg,\\
Luruper Chaussee 149, 22761 Hamburg, Germany\\
M.G.Ryskin\footnote{Work supported by the Grant INTAS-95-311, 
by the Russian Fund of Fundamental Research 98-02-17629,
and by Volkswagen Stiftung.}\\
Petersburg Nucl. Phys. Inst., Gatchina, S.Petersburg, 188350,
Russia\\[1cm]
\end{center}
\begin{abstract}
\noindent
The nonsinglet spin dependent structure function $g_2(x,Q^2)$ is studied
at small x within the double logarithmic  approximation  of 
perturbative QCD. Both positive and negative signature contributions
are considered, and we find a power-like growth in $1/x$. 
We discuss how our result fits into the Wandzura-Wilczek relation.
\end{abstract}

\section{Introduction}

The $Q^2$ evolution of the spin dependent structure function $g_1$ is
well known. In order to follow the evolution at $x\sim O(1)$ one 
can use the original Altarelli-Parisi
equation\cite{AP} or the anomalous dimensions calculated
in\cite{AR,MN}.
In the region of very small x it is possible to 
use the double logarithmic  approximation \cite{BER,BERs} to examine
the small-x behaviour of $g_1$.  In contrast to this, the situation with 
other polarized structure functions, in particular $g_2$, is not so clear 
at the moment.

The problem comes from the fact that $g_2$ has no parton
interpretation. $g_2$ describes the scattering on a transversely
polarized nucleon, and in terms of the Wilson operator product
expansion (OPE) it corresponds to the off-\-diagonal element of
the density matrix $\langle N(\lambda')|\hat
O|N(\lambda)\rangle$ with the helicities $\lambda'\neq\lambda$.

Let us recapitulate a number of well known facts about the relation between
the functions $g_1$ and $g_2$ (see, for example, the
review\cite{AEL}): \\ 
i) in the lowest approximation for the deep inelastic scattering (DIS)
on a free (on mass shell) quark, the function $g_2(x,Q^2)=0$ vanishes
identically ( see\cite{AEL} for a review); \\
ii) in the limit $m^2\ll Q^2$, the 
"massless" quark may have only the longitudinal polarization; so
 $g_1(x,Q^2)+g_2(x,Q^2)=0$ \cite{IKL}; 
\\
iii) the Wandzura-Wilczek relation\cite{WW} states, for the contribution of
twist-2 operators, that
\begin{equation}
g_1(x,Q^2)+g_2(x,Q^2)=\int\limits^1_x \frac{dy}y\ g_1(y,Q^2)\ ;
\end{equation}
iv) for the spin dependent structure function $g_2$ the next (twist-3) 
contribution is not suppressed, even for large
$Q^2$ \cite{HM}.

The presence of a twist-3 contribution $g_2^{(3)}(x,Q^2)$ in $g_2(x,Q^2)$ 
can be seen
in several ways. First of all, from an explicit one loop calculation 
\cite{ALNR,HZ} it is known that $g_2$ does not satisfy Eq.(1). 
Next, turning to the small-x region, it is again evident that the relation 
(1) cannot be valid for the whole functions $g_1$ and $g_2$: 
$g_1$ corresponds to an odd-signature amplitude, while $g_2$ consists of
even and odd signature pieces (see for example \cite{IKL,He}). 
Therefore $g_1$ and $g_2$ are expected to have quite 
different $x$-behaviour near $x=0$, i.e. they are proportional to 
different powers of $1/x$) at small $x$ (see \cite{BER,IKL,KL}).

In earlier papers \cite{AR,JK} a simple evolution equation for the moments of 
twist-3 parts of $g_2$ ($g_2^{(3)}(x,Q^2)$) function has been written, and the 
anomalous dimensions have been calculated for the scattering off a quark 
target. 
Subsequently, it has been realized \cite{SV} that the derivation in Ref. 
\cite{AR,JK} was incorrect, because quark operators and quark-gluon 
operators of the same twist and quantum numbers mix with each other. 
In \cite{BKL} a complete basis of operators 
has been suggested, and the mixing matrix has been calculated in \cite{BKL}. 
Similar calculations have later been perform by several other authors 
\cite{Ra,BB,JC}. Next in ref. \cite{KYTU} the anomalous dimensions were 
obtained using another gauge and renormalization scheme. The authors confirm 
the result of ref. \cite{BKL}, but they disagree with \cite{JC}.

The most significant departure from the standard leading-twist evolution 
equations 
\cite{AR,JK} is that the number of operators which contribute to the structure 
function $g_2^{(3)}(x,Q^2)$ is not fixed but increases with the moment index
$n$. Nevertheless, in the two limits:\\
i)number of colours $N_c\to \infty$, or\\ 
ii)the ($x$)-moment $n\to\infty$ \\
it has been shown that the quark-gluon operators decouple from the 
quark operator evolution equation \cite{ABH}, and the asymptotic behaviour of 
$g_2(x,Q^2)$ in the region $1-x\ll 1$ was derived.

In the region of small $x\ll1$ another new feature arises.
It is known that at small $x$ the strong ordering of
transverse momenta is violated \cite{GGFL,KL,EMR,BER}. 
Instead of $k^2_{i,t}\gg
k^2_{i-1,t}$ one has $k^2_{it}\gg\frac{x_i}{x_{i-1}}k^2_{i-1,t}$
(with $x_i\ll x_{i-1}$). As a result, double
log contributions of the form $(\alpha_s\ln^21/x)^n$ appear which
cannot be summed up in the framework of the conventional $\log
Q^2$ evolution. For $g_1$ this point was discussed in detail in
\cite{BER,BERs}. 

In this paper we study the small-$x$ behaviour of the nonsinglet polarized 
structure
function $g_2$, in very much the same way as in ~\cite{BER, BERs} $g_1$ has
been investigated. We will sum up all the leading $p$QCD double
logarithms (DL) of the form $\alpha_s^n\ln^k1/x\ln^mQ^2$ with
$m\le n$ and $k+m=2n-1$.

The outline of the paper is as follows. First in Sect.2
we will consider the non-singlet part of $g_2$ in the first loop
leading log approximation (LLA) and show that for the  massive
quark target (or for the $\mu$-meson in the QED case) in the
small x region $g^{(1)}_2=0$. Section 3 contains a detailed dicsussion 
of the two loop DL contribution, and in section 4 we derive the 
DL evolution of $g_2$ at small $x$. In section 5 we analyse the signature
content of $g_2$, and in section 6 we discuss the connection of 
our results with the Wandzura-Wilczek relation (1). In the final section we
present a brief summary.

\section{Definitions and The One Loop Leading Logarithm}

We start with a few definitions and general remarks.
The spin dependent antisymmetric part of the hadronic tensor
$iW^A_{\mu\nu}$ of the DIS lepton-hadron amplitude has the form:
\beqn
T^A_{\mu\nu}\ = \frac M{(pQ)}\ i \epsilon^{\mu\nu\alpha\beta}
\left[Q_\alpha  s_\beta T_1(x,Q^2)+Q_\alpha\left( s_\beta
-\frac{(sQ)}{(pQ)}p_\beta\right) T_2(x,Q^2) \right]\ ,
\eeqn
where $M$, $p_\mu$, and $s_\mu$ denote the mass, the four momentum and the 
spin vector of the target, resp. As usual, $(sp)=0$ and $s^2=-1$, and the
spin-dependent structure functions are defined as the discontinuities of
$T_1$ and $T_2$:
\beqn
g_{1,2}=-\frac{1}{2\pi} Im T_{1,2}.
\eeqn
Throughout our paper we work in the Feynman gauge.
Note that with the help of the identity ~\cite{He}
\begin{equation}
\epsilon_{\mu\nu\alpha\beta} p^\alpha s^\beta=
 \frac{\epsilon_{\mu\nu\alpha\beta}}{2x(pQ)}\left[(Qs)Q^\alpha
 p^\beta-(pQ)Q^\alpha s^\beta\right]
-(Q_\mu\epsilon_{\nu\alpha\beta\gamma}
 -Q_\nu\epsilon_{\mu\alpha\beta\gamma})\frac{p^\alpha Q^\beta
 s^\gamma}{Q^2}\ 
 \end{equation}
(where the last two terms may be dropped due to the gauge
invariance with respect to the photon) in (2) one can replace the 
vector in front of $T_2$ by $-2xp_\alpha s_\beta$. Then
the spin structure corresponding  to the function $g_2$ takes
the form  $(2xM)/(pQ)\epsilon^{\mu\nu\alpha\beta}$ $p_\alpha
s_\beta g_2$. It is this form which appears naturally when
we calculate the trace of the scattering amplitude.
\begin{figure}
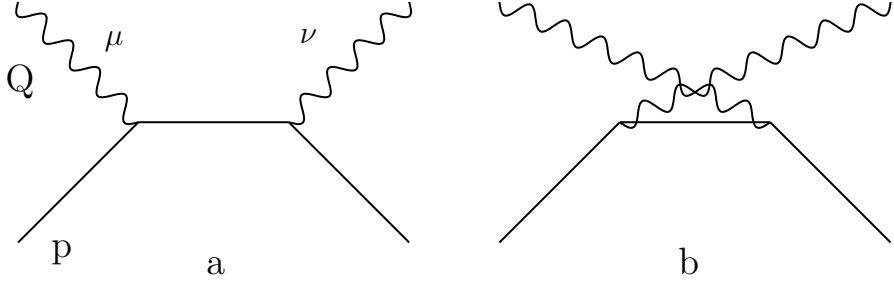

\begin{center}
\input paper56.fig1.pstex_t
\caption{The two Born diagrams}
\end{center}
\end{figure}

An important point that we have to discuss is the crossing symmetry of
the amplitude. Beginning with the Born approximation we have to sum the two 
Feynman diagrams Fig.1a and Fig.1b; they will be referred to as the
s-channel and the u-channel graphs. The
contribution of the u-channel graph may be obtained from the s-channel 
diagram just by permuting the photon indices
$\mu\rightleftharpoons\nu$ and by replacing $Q\to-Q$. The
whole amplitude, given by the sum of Fig.1a and Fig.1b
contributions, should be symmetric under this combined
$(\mu\rightleftharpoons\nu$, $Q\to-Q$) 
transformation (this reflects the Bose statistics of two
identical photons in $t$-channel).
To provide such a property the function $T_1$ should be {\em
odd}, i.e. it must have negative signature. On the other hand, the
crossing symmetry of the function $T_2$, which is multiplied by the factor
$(2xM)/(pQ)\cdot\epsilon^{\mu\nu\alpha\beta} p_\alpha s_\beta$
is more complicated and, in fact, cannot simply be read off from (2).
To illustrate the difficulty we simply note that, from the first part of
the second term in (2) (the piece proportional to $Q_{\alpha}s_{\beta}$), 
we would 
conclude that $T_2$ has the same signature as $T_1$. If, on the other hand,
we make use of the identity (4) and neglect (by gauge invariance) the second
term on the right hand side, the spin structure of $T_2$ in (2) can be 
replaced by $-2xp_\alpha s_\beta$. Now the same argument would lead to the
conclusion that $T_2$ has to be {\em even}, i.e. $T_2$ has positive 
signature. In section 5 we will show that $T_2$ has no definite signature
but can be written as a sum of two pieces with opposite signature.

To be definite let us consider the high energy scattering  
of a virtual photon on a heavy fermion with mass $m$. In order to obtain 
$g_1$ and $g_2$
we are interested in the energy discontinuity. So in our calculations we will
deal with the energy dicscontinuity, but in order to analyse the signature
properties we will have to come back to the full amplitudes $T_1$ and
$T_2$.
 
From the tensor structure of (2) one sees that we have to consider
a transversely polarized target. Namely, for a longitudinal polarization 
vector 
$s_{\mu}=p_{\mu}$ the vector in front of $T_2$ vanishes, and only $T_1$ 
contributes. For a transverse polarization, on the other hand, both
terms in (2) contribute, i.e. the result of such a calculation will
be proportional to $T_1+T_2$. We will therefore calculate the energy 
discontinuity of the scattering 
amplitude of the transverse polarized target which we will denote by
$g_{\perp}$. At the end we then subtract from the result
the (already known ~\cite{BER}) function $g_1$.  

To begin with the Born approximation, we only mention that in the tree 
approximation the amplitude contributes only to the first term
on the r.h.s. of (2). Therefore, the 
Born value of $g_{2}$, $g_{2\,\, Born}$, vanishes.  Contributions to the 
second term will appear only after the $\alpha_s$
corrections are taken into account.
\begin{figure}
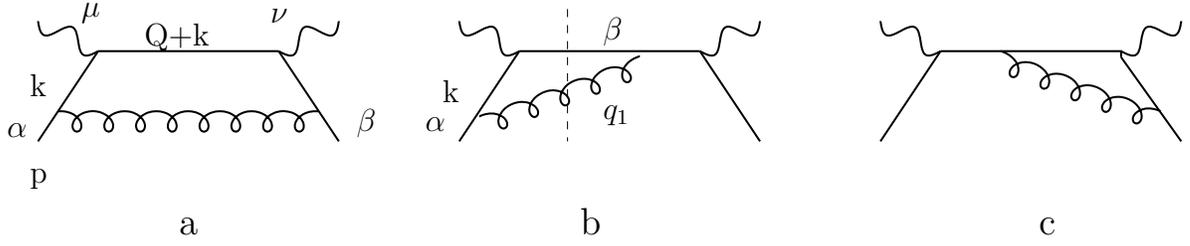

\begin{center}
\input paper56.fig2.pstex_t
\caption{The one-loop diagrams}
\end{center}
\end{figure}

To study the one loop contributions it
is convenient to use Sudakov \cite{Su} variables
\begin{eqnarray}
&& k_\mu=xp'_\mu+\alpha Q'_\mu+k_{t\mu};\;\; d^4k=\frac s2
 dxd\alpha d^2k_t;\;\; Q'^2=p'^2=0\ ; \nonumber \\
&& 2p'Q'=s \ ;\ \;\; Q_\mu=Q'_\mu -xp'_\mu\ ,
\end{eqnarray}
and to separate the transverse
and  longitudinal parts of the gluon propagator. In the Feynman
gauge the spin part of gluon propagator is:
\begin{equation}
d_{\mu\nu}\ =\ g_{\mu\nu} =g^\perp_{\mu\nu} +\frac{p'_\mu
Q'_\nu+Q'_\mu p'_\nu}{(p'Q')}\ .
\end{equation}
For the case of $g_2$  it is better to deal with the transverse
polarization of initial quark, as for the longitudinal quark
 only the function $g_1$ gives the leading contribution to
 the amplitude (2).

So the density matrix of the target quark is:
\begin{equation}
 \frac 12 (\hat{p}+m)(1-\gamma_5\hat{s}^\perp),
\end{equation}
and the spin dependent part of the trace in Fig.2a takes the form:
\begin{equation}
Tr^a=\ -\frac12 d_{\alpha\beta} Tr\left[\gamma_\alpha(\hat
p+m)\gamma_5\hat s^\perp\gamma_\beta(\hat k+m)\gamma_\nu(\hat
Q+\hat k+m)\gamma_\mu(\hat k+m)\right].
\end{equation}
Similarly, for the Fig.2b we obtain
\begin{equation}
Tr^b=\ -\frac12d_{\alpha\beta}Tr\left[\gamma_\alpha(\hat
p+m)\gamma_5\hat s^\perp\gamma_\nu(\hat Q+\hat p+m)\gamma_\beta
(\hat Q+\hat k+m)\gamma_\mu(\hat k+m)\right],
\end{equation}
where the quark momentum is $q_1=Q+p$.

It is easy to check that the transverse part of the gluon propagator
$d_{\alpha\beta}=g^\perp_{\alpha\beta}$ gives a vanishing leading log
contribution to Eqs.(8) and (9). Indeed, due to the identity
\begin{equation}
\sum_{\alpha\beta}g^\perp_{\alpha\beta} \gamma_\alpha \hat
s^\perp\gamma_\beta\ =\ 0
\end{equation}
the expression (8) equals to zero. The situation with Eq.(9) is a 
bit more complicated.

Note that as we deal with $s_\mu=s^\perp_\mu$ the 
polarization vectors of the photons ($\mu,\nu$) in (4) have to be different. 
One has to be longitudinal, whereas the other one is transverse and 
orthogonal to
$s^\perp_\mu$. If we chose $\gamma_\mu=\gamma^\perp_\mu$, 
$\gamma_\nu=\gamma^{\|}_\nu$ we find, according to (10), that $Tr^b=0$.
If, on the other hand, $\gamma_\nu=\gamma^\perp_\nu$ and 
$\gamma_\mu=\gamma^{\|}_\mu$ we arrive at:
\begin{equation}
Tr^b\ =\ -Tr\left[(\hat p-m)\gamma_5\hat s^\perp\gamma^\perp_\nu
(\hat Q+\hat p-m)(\hat Q+\hat p-m)(\hat Q+\hat
k+m)\gamma_\mu(\hat k+m)\right]\ .
\end{equation}
At small $x$, in the quark propagator we simply put $1/k^2\simeq1/k^2_t$.

In order to obtain the leading logarithm $dk^2_t/k^2_t$ we have
to keep in (11) only the longitudinal $(k_\mu\simeq xp)$ part of
quark momentum and, due to the smallness of the quark mass $m^2\ll
k^2_t$, one may keep just the first power of $m$. Then Eq.(11)
takes the form
\begin{equation}
 Tr^b\ \simeq\ -(1-x)mTr\left[\hat p\gamma_5s^\perp\gamma^\perp_
\nu\hat Q(\hat Q+\hat k)\gamma^{\|}_\mu\right]\ =\ 0 \ ,
\end{equation}
as in our LLA limit $Q+k\simeq Q+xp=Q'$ and $\hat p\hat Q\hat
Q'=-xp^2\hat Q'+\hat pQ'^2=-xm^2\hat Q'\simeq0$; recall that we
neglect the contributions of the order of $m^2$.

Now let us consider the longitudinal part of the gluon propagator,
i.e. the second term of (6). Keeping only the lowest power of mass the mass m,
and taking into account that
$$ d^{\|}_{\alpha\beta} \gamma_\alpha(\hat p+m)\gamma_5 \hat,
s^\perp \gamma_\beta\ =\ 2m\gamma_5 \hat s^\perp  $$
we obtain for the diagram shown in Fig.2a:
\begin{eqnarray}
Tr^a&=&-mTr\left[\gamma_5\hat s^\perp\hat k\gamma_\nu(\hat
Q+\hat k)\gamma_\mu\hat k\right]= -mk^2Tr\left[\gamma_5\hat
s^\perp\gamma_\nu(\hat Q+\hat k)\gamma_\mu\right] \nonumber\\
&& -\ m\,2(ks^\perp)Tr\left[\gamma_5\gamma_\nu(\hat Q+\hat
k)\gamma_\mu\hat k\right]\ .
\end{eqnarray}
Here one has two quark propagators $(1/k^2)^2\approx1/k^4_t$. So in order 
to obtain the leading logarithm we have to keep in the numerator
the terms proportional to $k^2_t$. After the averaging over the
azimuthal angle of the vector $\vec{k}_t$ the last term in (13) may
be written as
$$ 2m(ks^\perp)Tr\left[\gamma_5\gamma_\nu(\hat Q+\hat
k)\gamma_\mu\hat k\right]=mk^2Tr\left[\gamma_5\gamma_\nu(\hat
Q+\hat k)\gamma_\mu\hat
s^\perp\right]+mk^2Tr\left[\gamma_5\gamma_\nu\hat
s^\perp\gamma_\mu\hat k\right].$$
Now the first term cancels against the corresponding term in (13). In
the last term, within the LLA, we may put $\hat k=x\hat p$, and
finally obtain
\begin{equation}
Tr^a\ =\ +mxk^2Tr\left[\gamma_5\hat
s^\perp\gamma_\nu\gamma_\mu\hat p\right]\ .
\end{equation}

For the case of Fig.2b the first longitudinal term
$d_{\alpha\beta}=p'_\alpha Q'_\beta/(p'Q')$ gives zero (within
 the LLA).  Thus we have
\begin{equation} Tr^b=\
 -\frac1{2(p'Q')}Tr\left[\hat Q'(\hat p+m)\gamma_5\hat
 s^\perp(\hat Q+\hat p+m)\hat p'(\hat Q+\hat k+m)\gamma_\mu(\hat
k+m)\right]\ .
\end{equation}
As before, in order to obtain the leading logarithm
we put $\hat k=x\hat p$.  Next it is useful to note
that $(\hat Q+\hat p+m)\hat p'(\hat Q+\hat k+m)=2(p'Q')(\hat
Q'+m)$. Therefore
\begin{eqnarray} && \hspace{-1.cm} Tr^b=-Tr\left[\hat Q'(\hat
p+m)\gamma_5\hat s^\perp\gamma_\nu \hat
Q'\gamma_\mu(xp+m)\right]-mxTr\left[\hat Q'\hat p\gamma_5 \hat
s^\perp\gamma_\nu\gamma_\mu\hat p\right] \\
 && \hspace{-1.cm}
=-mTr\left[\hat Q'\hat p\gamma_5\hat s^\perp\gamma_\nu\hat
Q'\gamma_\mu\right]-(2pQ)mxTr\left[\gamma_5\hat
s^\perp\gamma_\nu\gamma_\mu p\right]-mTr\left[Q'\gamma_5s^\perp
\gamma_\nu Q'\gamma_\mu p\right]x.     \nonumber
\end{eqnarray}
The first term on the r.h.s. of (16) is equal to zero if
$\gamma_\mu=\gamma^\perp_\mu$. If on the other hand we take 
$\gamma_\mu=\gamma^{\|}_\mu$ $(\gamma_\nu=\gamma^\perp_\nu)$ we find
$$ -Tr\left[\hat Q'\hat p\gamma_5
s^\perp\gamma^\perp_\nu Q'\gamma_\mu\right]\ =\ 2(pQ)Tr\left[
\gamma_5s^\perp\gamma^\perp_\nu\gamma^{\|}_\mu Q'\right]\ =\
-x(2pQ)Tr\left[\gamma_5s^\perp\gamma^\perp_\nu\gamma^{\|}_\mu
p\right]. $$
To obtain the last equality we have used the fact that the
longitudinal photon polarization vector $e^{\|}_\mu=
\frac1{\sqrt{Q^2}}(Q'+xp')_\mu$, i.e.
$\gamma^{\|}_\mu\|(Q'+xp')_\mu$. Therefore in the last
expression one can replace $\hat Q'$ by $-x\hat p$.
The factor $2pQ$ in Tr$^b$ (15) cancels the rightmost quark
propagator $1/(Q+p)^2\simeq1/(2pQ)$. 

Finally, we have to add the
contribution of the diagram 2c. After combining it with our results for 
Figs.2a and b the whole one-loop LLA result takes the form
\begin{equation}
\frac{im}{(pQ)}\,\epsilon^{\mu\nu\alpha\beta}s^\perp_\alpha
Q_\beta\cdot(g_1+g_2)=\frac{e^2_qC_F\alpha_s}{8\pi(pQ)}\,
m\int\frac{dk^2_t}{k^2_t}\, Tr\left[\gamma_5\hat s^\perp
\gamma_\mu\gamma_\nu\hat Q\right][(1+O(x)],
\end{equation}
where $e_q$ is the electric charge of the quark $q$,
$C_F=(N^2_c-1)/2N_c$, and $\alpha_s$ is the QCD coupling
constant.

The well known result for the non-singlet function $g_1(x,Q^2)$
in the one loop approximation and LLA small $x$ limit is::
\begin{equation}
g_1\ =\ \frac{e^2_q C_F\alpha_s}{2\pi} \int
\frac{dk^2_t}{k^2_t}\ .
\end{equation}
Comparing  Eqs.(17) and (18) we see that in this limit
$g_2=0+O(x)$, in agreement with the more precise computation
\cite{ALNR}
\begin{equation}
g_2(x,Q^2)\ \simeq\ \frac{e^2_qC_F\alpha_s}{2\pi}\ x\ln
\frac{Q^2}{m^2x(1-x)}\ .
\end{equation}
\begin{figure}
\begin{center}
\input paper56.fig3ac.pstex_t
\\[0.6cm]
\end{center}
\end{figure}
\begin{figure}
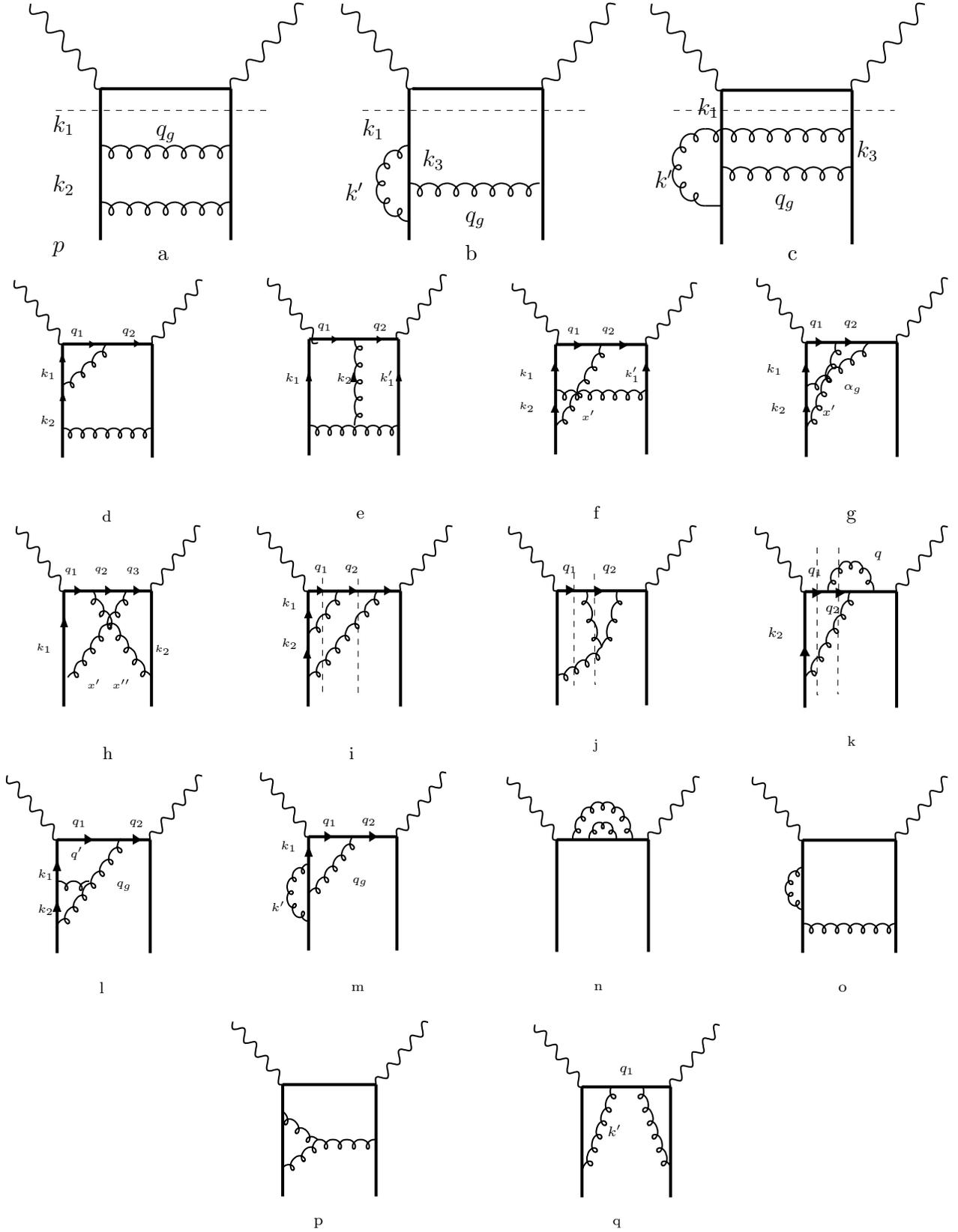

\begin{center}
\input paper56.fig3dq.pstex_t
\caption{The two-loop diagrams}
\end{center}
\end{figure}

\section{Second loop}

Next we turn to the two loop contributions to $g_2$. The diagrams are 
shown in Fig.3. In this section we will show that only the first 
three diagrams in Figs.3 a -c lead to a double logarithms, whereas in all 
other diagrams the double logarithm cancels. 
These cancellations are mainly due to the AFS-mechanism ~\cite{AFS,Mst}: 
first one realizes that, in the approximation which leads to the double  
logarithms, it is enough to keep the 
nonzero value of the $x$ variable only in those propagators which have the 
largest $\alpha$. We then find 
for the integration of either $x_1$ or $x_2$
that the propagator poles lie on the same side of the integration contour.
Their sum adds up to zero, since closing the contour in the opposite half plane
(which in this approximation has no propagator poles) gives zero.

To see this in detail, let us go through the diagrams in Fig.3. First we 
eliminate all self energy insertions to Fig.2a or b: they never change
the spin structure and, in the Feynman gauge, the self energy loop 
has only a single logarithmic divergence. Examples are shown in Figs.3n and
3o.

Next we consider the vertex correction in Fig.3d. The absence of the double
logarithm in the last step can be checked
by a straightforward computation of the trace (analogous to (9),(15)
for the case of Fig.2b). Alternatively, one can use the "AFS" type arguments
and consider the cancellation between the cuts passing through
the quark lines with $q_1$ or $q_2$.

Next we consider the graphs in Figs.3e-h. In order
to obtain the double log contribution we have to consider
the domain where all the momenta are strongly ordered. Say, in
Fig. 3g $1\gg x_2\gg x_1$, $k_{t2}\ll k_{t1}$ and
$|\alpha_1|\gg|\alpha_2|$. Let us consider the loop with the
largest values of the $\alpha$ component (the photon momentum fraction).
This loop includes 3 quark propagators $(q_1,q_2$ and $k_1$) and
one gluon $(\alpha_g=\alpha_1-\alpha_2)$. Two propagators with
the largest $\alpha_{q_1}\simeq\alpha_{q_2}\simeq1$ are enough
to provide the convergency of the integral over  $dx_1$. Thus we
may neglect all other propagators (with the order of $\alpha_1$
accuracy). For the $1/q^2_1$ and $1/q^2_2$ propagators
both poles lie in the same semiplane, and closing the $x_1$
contour in the opposite semiplane one gets zero result. In other words, the
contributions of the $1/q^2_1$ and $1/q_2^2$ poles cancel each
other. This is analogous to the famous cancellation of the AFS
\cite{AFS} pomeron cut contribution which was first discussed by
Mandelstam \cite{Mst}. This cancellation reflects the fact that
the diagram shown in Fig.3g has a wrong space-time structure. In
the Breit (photon rest) frame the formation time of the gluon with
momentum $q'=p-k_2$, $\tau'\simeq1/(\alpha'\sqrt{Q^2})$, which is 
proportional to the inverse power
of the $\alpha'\simeq\alpha_2$ component
$\tau'\simeq1/(\alpha'\sqrt{Q^2})$ is much larger than the
formation time of the second gluon with $q=k_2-k_1$,
$(\tau\simeq1/(\alpha_1\sqrt{Q^2})$, and also much larger than the
$s$-channel quark time life, $\tau_q\sim1/\sqrt{Q^2})$ (the separation 
between the points $\mu$ and $\nu$ in Fig.1). On the other
hand, along the upper quark line in Fig.3g we have quite the opposite,
wrong time ordering $\tau_q>\tau>\tau'$.
\begin{figure}
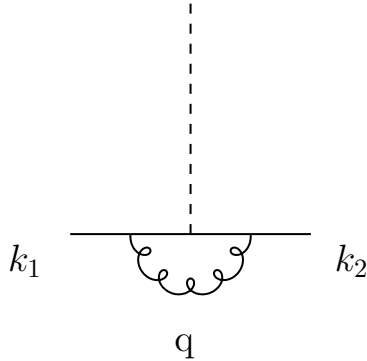

\begin{center}
\input paper56.fig4.pstex_t
\caption{Structure of the Sudakov vertex}
\end{center}
\end{figure}
By the same reason the graphs of Fig.3 e,f graphs give no leading log
contribution. In Fig.3e we have 
$-\alpha_1=|\alpha_{k_1}|\gg\alpha'_1$. Then in the right hand side loop
the two propagators with the largest $\alpha$-components are
$q_2$ ($\alpha_q\sim1$) and $k_2=q_2-q_1$ ($\alpha_2\sim\alpha_1)$.
In both propagators the photon momentum fraction $\alpha$
goes in the same direction, i.e. we have both poles in the
same $x_2$-semiplane, and closing the contour of the $x_2$
integration in the opposite semiplane one obtains zero.
In the case of large $\alpha'_1\gg-\alpha_1\quad$ (of
course, we still have $\alpha'_1\ll1$ and $-\alpha_1\ll1$) we
obtain the same  zero result, due to the $x_1$-integration
in the left hand side loop.
Finally, for the diagrams Fig.3f and h we consider the upper
$(k_1,q_1,q_2,k'_1)$ loop (without the quark $k_2$): the two
propagators with the largest photon momentum fraction $\alpha$
are $q_1$ and $q_2$, and we have the same cancellation as for the
case of Fig.3g.

The diagrams in Figs.3i-k contain vertex corrections as subgraphs.
The general structure of such a `Sudakov` vertex is illustrated in Fig.4. 
The double logarithmic (DL) contribution has been analysed first in \cite{Su}.
It comes from the
region of $\alpha_1\gg\alpha_q\gg\alpha_2$, $x_1\ll x_q\ll x_2$
(for the case of $\alpha_1\gg\alpha_2$, $x_1\ll x_2)$. The
corresponding kinematical domain for the graphs of Fig.3 i,j,k 
is $1\gg\alpha_q\gg\alpha_2$, with  $q=k_2-k_1$ in Fig.3i and
$q=q_1-q_2$ in Fig.3j,k. Indeed, say, in the upper triangle of
Fig.3i one may close the $x_1$-contour around the $1/q^2$ pole of the 
gluon propagator. Then the quark propagators
$1/k^2_1\approx1/(x_2\alpha_qs)$ (here $x_1=x_2-x_q\approx x_2$
as $x_q\ll x_2$) and $1/q^2_1\simeq1/x_qs$. To obtain the
leading logarithm one has to consider the longitudinal
polarization of gluons $d^{\|}_{\alpha\beta}=Q'_\alpha
p'_\beta/(Q'p')$ (as we have seen in Sec.2). In the trace
corresponding to the quark loop it gives $\hat Q'\hat k_2\hat
Q'=x_2s\cdot\hat Q'$ and $\hat p'\hat q_2\hat p'=s\cdot\hat p'$.
In other words, we reproduce the original spin structure of
the diagram Fig.2b and obtain the factor $x_2$ which cancels the
$1/x_2$ coming from the $1/k^2_1$ propagator.
Thus the loop integral
\begin{equation}
\int\frac{dk^2_{1t}d\alpha_qdx_q}{\alpha_q\ x_q}\
\delta\left(\alpha_q x_q s+(k_1-k_2)^2_t\right)\
=\int\frac{d\alpha_1}{\alpha_q}\ \frac{dx_q}{x_q}
\end{equation}
takes the double  logarithmic form.
However in such a domain for the second loop we have the
"AFS"-type cancellation again. In Fig.3i for the loop with 2
gluons and quarks $k_2$ and $q_2$ the two largest $\alpha$
($\alpha_q$ and $\alpha_{q_2}\simeq1$) go in the same direction, leading to 
two poles in the same $x_2$ semiplane. In the left loop of Fig.3j
(which does not include the quark $q_2$) the two largest $\alpha$ variables are
$\alpha_{q_1}$ and $\alpha_q$ which flow in the same direction
as well. For the case of Fig.3k one has the same cancellation as
in Fig.3g. In the lower triangle (with 3 quark and one gluon
propagators) the two largest $\alpha$ are $\alpha_{q_1}\simeq1$
and $\alpha_{q_2}\simeq1$.
Here (in the vertex correction) the absence of a DL contribution
may be treated as a well known cancellation of the leading logs
between the virtual (Sudakov formfactor) and real "soft" gluon
emission in the inclusive cross section; i.e. the cancellation
between the cuts shown by the left- and right-dotted lines in
Fig.3 i,j,k.

For the case of the vertices shown in Fig.3l, it is more convenient to
fix the value of $x'$ within the conventional DL region $(1\gg
x'\gg x_1)$ and then to observe the "AFS"-type cancellation in
the upper loop $(k_1,q_1,q_g$, but without the $k'$-propagator)
due to the two largest fraction of the target momentum $x_{q'}
=x'-x_1$ and $x_g\approx1(q_g=q_2-q_1=x_gp+\alpha_gQ'+q_t$) which
go in the same direction.\\

The vertex Fig.3m does not give the double log at all.  For $x'\ll 1$ and 
$\alpha '\ll\alpha_1$ the virtuality $(k'-k_1)^2$ of the quark propagator 
is controlled not by the value of $\alpha_1x's$ but by $k^2_{1t}\sim\alpha_1s$
\footnote{If instead of the gluon $q_g$ one cuts the quark lines $(p-k')$ and 
$(k_1-k')$ (as in Fig.5a) we loose the logarithm in the upper loop (with gluon 
$q_g$). Indeed, the quarks  $(p-k')$ and $(k_1-k')$ are now on mass shell, and 
thanks to the gauge invariance one may replace the polarization vector $Q'$ 
 (at the lower end of the gluon $q_g$) by $-q_{gt}/{\alpha_g}\simeq 
-k_{1t}/\alpha_1$. This factor $(k_{1t}/\alpha_1)$ kills the normal Dlog in the
 loop.}.
Finally, in the case of the two photon vertex corrections (Fig.3q) 
one has to "cut" 
one of the vertices in order to have a small $x$. Then the central quark 
virtuality is $q^2_1\simeq s$, and again we loose the double logarithm in the 
second vertex, since the virtuality $(q_1-k')^2\simeq s$ of the next quark 
propagator is not controled by the gluon loop ($\alpha' ,x'$) variables.
The same is true for the diagrams in Figs.3j and 3r. 
\begin{figure}
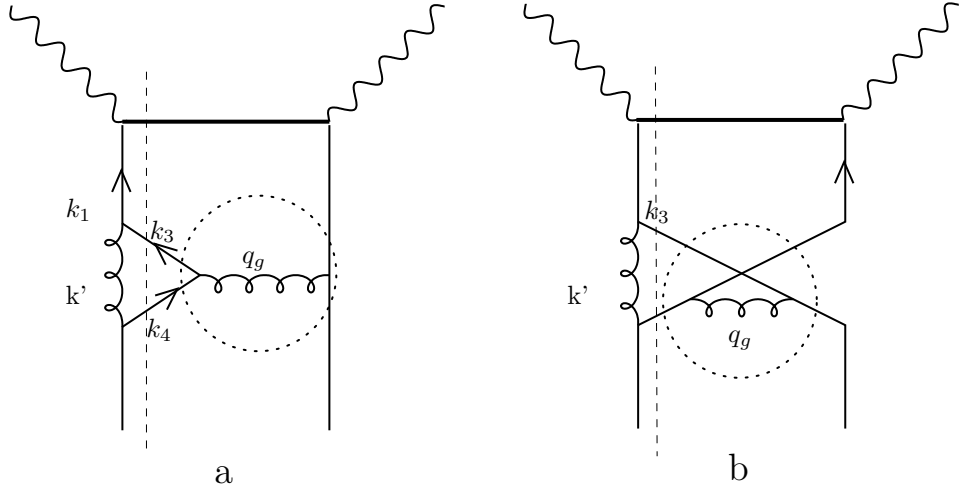

\begin{center}
\input paper56.fig5.pstex_t
\caption{Illustration of the Signature properties of Figs.3b and c}
\end{center}
\end{figure}

Finally we return to the `reducible` graphs of Fig.3a-c. They all have the
property that they can be divided into two parts by cutting two t-channel
quark lines. First of all we will check
the cancellation of the double logarithms corresponding to the
lower (below the crosses) part of the graphs in Fig.3b and c, i.e. the
cancellation at the level of the quark-quark
scattering amplitude. Such a cancellation was first
observed in QED \cite{GGFL} and was then discussed in \cite{KL}
for the DL quark-quark amplitude in QCD.

It is known that the DL contribution comes from the longitudinal
$d^{\|}_{\alpha\beta}$ spin part of the gluon $k'$
propagator in  the region of $x\ll x'\ll1$ and
$|\alpha'|\ll|\alpha_1|$. The contour of the $x'$ integration may
be closed around the quark $1/k^2_3$ pole, and then the contour of
the next $x_g$ integral (see Fig.5a,b) should be closed around the
$1/k^2_4$ pole (or one has to use the $1/(k^2_4+i\epsilon)$
propagator to take the imaginary part of the whole amplitude).
This gives the ''Sudakov''-type double logarithm
$$ \int\frac{d\alpha'}{\alpha'}\ \frac{dk'^2}{k'^2_t}\ \quad
\mbox{ or }\ \quad \int\frac{d\alpha'}{\alpha'}\ \frac{dx'}{x'},
$$
similar to Eq.(20), but using the $\delta$-function
$\delta(k^2_3)$ with $k^2_3\approx\alpha_1x's+k^2_{3t}$ for
the $d^2k'_t$ integral $(\vec{k}_{3t}=\vec{k'}_{t}-\vec{k}_{1t}$) rather than 
the $sdx'$ integral. In terms of the $s$-channel intermediate states
it means that we consider the cut of Fig.3b,c diagrams shown in
Fig.5a,b by the dash-dotted line; the upper part of Fig.3b,c
is shown in Fig.5 by dashed lines. Note that in Fig.5 we have
crossed the upper quark line in order to plot both the
graphs in the same form. The even signature (corresponding to
$g_2$ structure function) amplitude is invariant under the  crossing 
transformation, and the only
difference between Fig.5a and 5b (or 3b and 3c) comes
from the colour charge of the $k_3$ quark line. The two t-channel
quarks form a singlet colourless state and
thus should have an opposite colour sign. Therefore, within the
DL accuracy Fig.3b and 3c, diagrams cancel each other.

Note that such cancellation takes place not only in the case of 
Fig.3b,c but for any DL quark-quark amplitude
(in Fig.5 the quark-quark amplitude is denoted by the dotted circle). 
This cancellation, i.e. the absence of the non-ladder
graphs in the DL even signature quark-quark QCD amplitude was
proven and discussed in more detail in \cite{KL}.

As a result of all these cancellations, the only DL two loop
contribution to the even signature amplitude comes from the simple 
diagram in Fig.3a with
the two-quark $t$-channel intermediate state. However, the
calculation of this pure ladder graph Fig.3a is not so trivial, as we
have to take care of the longitudinal parts of the quark
momenta $k_1$ and $k_2$. For example, even in the DL kinematics
the product of 4 vectors
$2(k_1k_2)=k^2_1+k^2_2-q^2_g=k^2_1+k^2_2\neq2(k_{1t}\cdot k_{2t})$;
here we put the gluon $q^2_g=0$ on mass shell.
That is why we will describe the algorithm of the computation in
the next section.
\begin{figure}
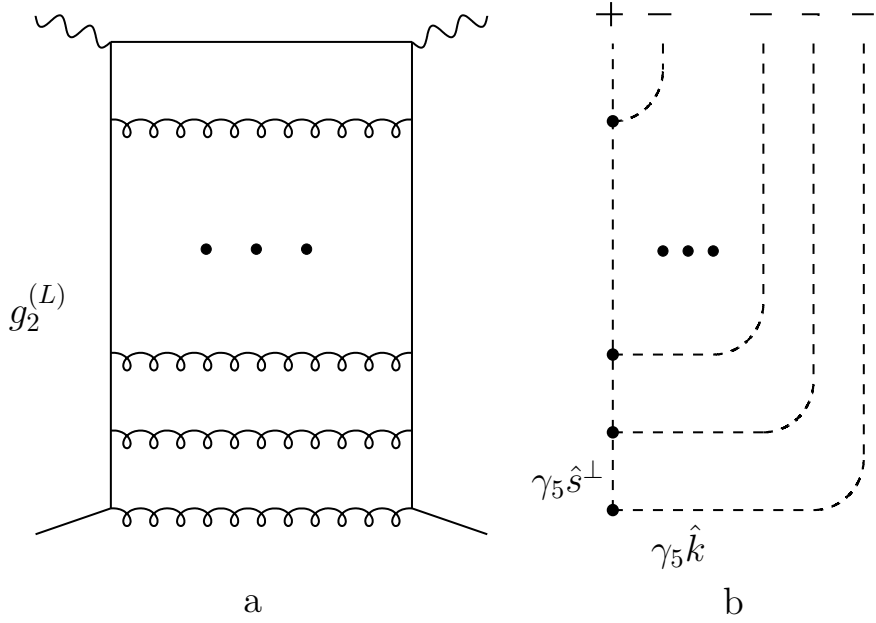

\begin{center}
\input paper56.fig6.pstex_t
\caption{Branches in the ladder graphs}
\end{center}
\end{figure}

\section{Evolution of Ladder Graphs}

As we have shown, in the two loop approximation only the diagrams
Fig.3a-c give double logarithms. In this section we
generalize our calculation of the ladder graph (Fig.3a) to all orders (Fig.6).
In the next section we will then return to the non-ladder graphs in Fig.3b,c. 

We begin at the bottom part of Fig.6a and work our way upwards.
After the first gluon rung the spin dependent part
of the density matrix (7) takes the form
\beqn
\sum_a\ \gamma_a\ \frac{\hat p+m}2\ \gamma_5 s^\perp\gamma_a\ =\
m\gamma_5\hat s^\perp\ .
\eeqn
As before we keep only the first power of mass $m \ll\sqrt{Q^2}$,
and therefore one may neglect the mass in all other propagators.
Then we write the evolution kernel as
\begin{equation}
\frac{C_F\alpha_s}{2\pi}\int\frac{dx}x\frac{dk^2_t}{k^4}\
\hat{\cal K}(\gamma_5\hat s)
\end{equation}
with
\begin{equation}
\hat{\cal K}(\gamma_5\hat s)\ =\ \hat k\gamma_5\hat s\hat k\ =\
(k^2)\gamma_5\hat s- 2(sk)\cdot\gamma_5\hat k\ .
\end{equation}
The first term on the r.h.s. of (23) repeats the initial spin
structure $(\gamma_5\hat s)$ while the second one has the new
matrix structure $\gamma_5\hat k$. When upper rungs are included, this
second term generates a ``new branch``
(Fig.6b). Let us first follow this new
branch. As there are only two external 
momenta ($Q$ and $k$) available for this new term, 
the subsequent DL evolution with $n-1$ rungs gives a contribution of the form
\beqn
 2(s^\perp k)\cdot \epsilon^{\mu\nu\alpha\beta} k_\alpha
Q_\beta \cdot f^{(n-1)}(k,Q)
\eeqn
where the tensor structure comes from the final trace, and the scalar 
function $f^{(n-1)}(k,Q)$ is
equal (in (n-1)th order $\alpha_s$) to the ladder part of the 
structure function $g_1(x',Q^2)$
(with $x'=Q^2/2(k\cdot Q))$. Since the first rung in (22) and (23) has the same
color factor $C_F\alpha_s/2\pi$ as the evolution kernel in the
ladder part of the longitudinal structure function (see below), we can absorb this rung
into $f^{(n-1)}$ and obtain $f^{(n)}$. Finally, from the integration over
the azimuthal direction of the vector $k_t$ the product
$2(s^\perp k)\epsilon^{\mu \nu\alpha\beta}k_\alpha
Q_\beta$ turns into $k^2_t\epsilon^{\mu\nu\alpha\beta}s^\perp_\alpha
Q_\beta$. In this way we obtain the factor $k_t^2$ which is needed for the
logarithmic integral $k^2_tdk^2_t/k^4$ in the lowest cell of the ladder.
As a result, after summation over n, the ``new branch`` resulting from the 
second term in (23)
leads to a contribution of the form ~\cite{BER}:
\beqn
g^L(x,Q^2)\ &=&\ \frac{e^2_q}2\int\limits^{i\infty}_{-i\infty}
\frac{d\omega}{2\pi i}\left( \frac1x \right)^\omega
\left(\frac{Q^2}{\mu^2}\right)^{f_0(\omega)/2}\frac\omega{\omega
-f_0(\omega)/2} \nonumber \\
& = & \frac{e^2_q}2 \delta(1-x)+\frac{e^2_q}2\cdot
\frac{C_F\alpha_s}{2\pi}\ln\frac s{\mu^2}+\sum^\infty_{n=2}(C_F
\alpha_s)^n g^{(n)}_L(x,Q^2)\ ,
\eeqn
where instead of the amplitude $f^{(-)}_0$ in \cite{BER} we now have
put the even signature quark-quark amplitude:
\begin{equation}
f_0(\omega)\ =\ \omega-\sqrt{\omega^2-\frac{2C_F}\pi\ \alpha_s}\ ,
\end{equation}
which describes the pure ladder DL evolution and does not contain the
nonladder graphs.
   
Let us now return to the ``normal`` branch in (23) (Fig.7) with the structure 
$\gamma_5 \hat{s}^{\perp}$. At the next-to-lowest rung we simply 
repeat the argument in (23): we again find, 
in addition to the `old` structure, the beginning of a new one branch which,
after multiplication with the two lower rungs and summation over n,
again leads to a contribution (25).
Moving up the ladder in Fig.6, this procedure repeats itself at each rung,
i.e. for each order $\alpha_s$ we obtain a contribution (25). Formally,
this multiplicity of (25) is achieved by taking the derivative
with respect to $\alpha_s$ and then multiplying with $\alpha_s$:
\beqn
 - \alpha_s \frac{\partial}{\partial \alpha_s} g^L
\eeqn
where the minus sign reflects the sign of the second term in (23).

Finally we have to complete the `normal` branch which is obtained by following,
at each rung, the tensor structure $\gamma_5 \hat{s}^{\perp}$.
As we have already said before, for each rung we collect the same coefficient
as in the case of the longitudinal target polarization \footnote{For the 
longitudinal
case the analogue of (23) is $\hat{k} \gamma_5 \hat{p} \hat{k} =
(k^2)\gamma_5 \hat{p} - 2(pk) \gamma_5 \hat{k}$. In the leading-log kinematics
we have $2(pk)=\alpha s \approx k_t^2$, and in the last term we have to keep
only the longitudinal part of $\hat k\simeq x\hat p$.
Thus one may neglect this last $(2(pk)\gamma_5\hat k\simeq
k^2_t\cdot x\gamma_5\hat p)$ term in comparison with the first
$(k^2\cdot\gamma_5\hat p)$ one, and there is no `new branch` in the 
longitudinal case.}. Having reached the upper end of the ladder,
we arrive at the "coefficient function" of the ``normal branch``
$m\gamma_5\hat s^\perp$, which results from taking the trace at the upper
end of the ladder.  This trace
$$ Tr\left[\gamma_5\hat s^\perp\gamma_\nu(Q+k)\gamma_\mu\right]\
\simeq\ Tr\left[\gamma_5\hat s^\perp\gamma_\nu(Q+xp)\gamma_\mu
\right] $$
is by a factor $1/2$ smaller than the trace $Tr[\gamma_5s^\perp\gamma_\nu
Q\gamma_\mu]$ in the case of the function $g_1$. The DLA result
for the (ladder part of) spin dependent non-singlet structure function
is given in (25). So the result for the ``normal branch`` is simply
$\frac{1}{2} g^L$.

Taking the sum of the all contributions, we arrive at our final result for
the ladder graphs in Fig.6:
\beqn
g_{\perp}^L=
\left( \frac{1}{2} - \frac{\partial}{\partial \ln \alpha_s} \right) g^L(x,Q^2)
\eeqn
Since we have studied the transverse polarization of the target, our
result represents the ladder-graph contribution to the sum $g_1+g_2$.
Before we subtract $g_1$, we first turn to the question of signature
and the nonladder graphs.

\section{Signature and the Non-Ladder Diagrams}
\begin{figure}
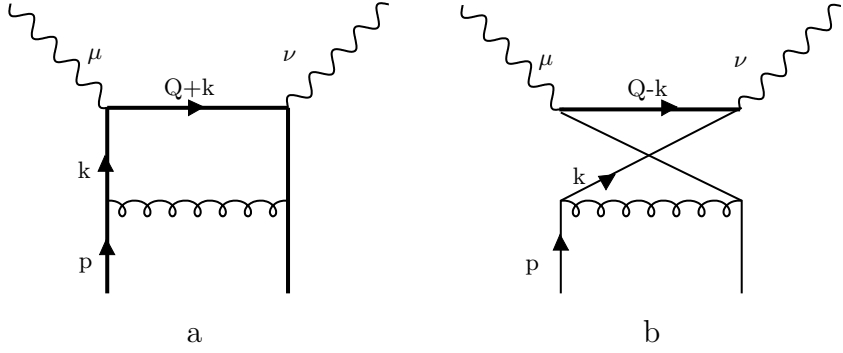

\begin{center}
\input paper56.fig7.pstex_t
\caption{Signature structure of the first loop}
\end{center}
\end{figure} 

So far we have generalized our two loop calculation only 
to the summation of ladder graphs. Now we have to consider the non-ladder 
graphs Figs.3b and c and discuss their generalization to higher orders.
In section 3 we have argued that the double logarithms of Figs.3b and
c cancel if we take the even signature combination of Figs.3b, c and their
$s-u$ crossed counterparts, but add up if we  
consider the odd signature combination (Fig.7). In this section we will 
analyse the signature content of the ladder graphs: we will find
that a general ladder graph $\cal F$ of Fig.6 contains both even and odd 
signature pieces. For the even signature contributions we then know that
the nonladder contributions cancel, and only the odd signature pieces
of the ladder graphs will receive extra contributions from the nonladder
graphs. The summation of the nonplanar diagrams is done
in the usual way of ~\cite{KL,BER,BERs}.

Let us therefore take a closer look at the signature properties of the
ladder graphs. For this it is useful to discuss the amplitudes rather than
the energy discontinuities. At the bottom of Fig.6 we have the density 
matrix (21),
$m\gamma_5 \hat{s}$. The contributions of Figs.7a and b to the amplitudes
$T_1$ and $T_2$ are proportional to:
\beqn
\frac{1}{2Qk+Q^2+k^2}\;Tr\left[ \hat{k} \gamma_5 \hat{s} \hat{k} \gamma_{\nu}
(\hat{Q} + \hat{k}) \gamma_{\mu} \right]
\eeqn
and
\beqn
\frac{1}{-2Qk+Q^2+k^2}\;Tr\left[ \hat{k} \gamma_5 \hat{s} \hat{k} \gamma_{\mu}
(-\hat{Q} + \hat{k}) \gamma_{\nu} \right].
\eeqn 
As before, we make use of eq.(23)
$\hat{k}\gamma_5\hat{s}\hat{k} =\
(k^2)\gamma_5\hat{s}- 2(sk)\gamma_5\hat{k}$ (which is valid for any spin 
vector $s_{\mu}$, not only for $s_{\mu}^{\perp}$). Concentrating on the 
double logarithmic region
$x_k\gg x_{bj}=Q^2/s$, in the denominators we retain only the $2Qk$ term.
The contributions (29) and (30) then take the form:
\beqn
\frac{1}{2Qk} \; Tr \left[ k^2 (\gamma_5 \hat{s} - 2(sk) \gamma_5 \hat{k})
           \gamma_{\nu} (\hat{Q} + \hat{k})\gamma_{\mu} \right]
\eeqn
and 
\beqn
- \frac{1}{2Qk} \; Tr \left[ k^2 (\gamma_5 \hat{s} - 2(sk) \gamma_5 \hat{k})
           \gamma_{\nu} (\hat{Q} - \hat{k})\gamma_{\mu} \right]
\eeqn
(in the last equation we have written the $\gamma$-matrices in the same order
as in (31)). 
One immediately sees that for the first piece ($\gamma_5\hat{s}$) two 
contributions with opposite signature
emerge: the first one (proportional to $Tr [\gamma_5 \hat{s} \gamma_{\nu} \hat{Q} \gamma_{\mu}]$) 
has odd signature, whereas the second one (proportional to
$Tr [\gamma_5 \hat{s} \gamma_{\nu} \hat{k} \gamma_{\mu}]$) has even 
signature. The second term ($\sim 2(sk) \gamma_5 \hat{k}$) again leads to two 
contributions, but the second one vanishes (it has two identical $\hat{k}$ 
vectors, and the trace with the $\gamma_5$ matrix vanishes). So we have only
$2(sk) Tr[ \gamma_5 \hat{k} \gamma_{\nu} \hat{Q}\gamma_{\mu}]$, and this 
structure has odd signature. Since this is the term which generates the
`new branch` we conclude that the `new branch` is purely odd, whereas the `old 
branch` contains both signatures.

Next we have to relate this decomposition to the tensor structures appearing 
in (2). Let us first demonstrate explicitly that $g_1$ has only odd 
signature (cf. the discussion after (4)). For this we briefly return to 
the longitudinal polarization of 
the target. In footnote 3 we have argued that for $s_{\mu}=p_{\mu}$ the
second term in the decomposition (23), $2(pk)\gamma_5\hat{k}$, can be 
neglected in comparison with the first one, $\gamma_5 \hat{p}$. In the 
remaining term $Tr[\gamma_5\hat{p}\gamma_{\nu} (\hat{Q}+\hat{k})\gamma_{\mu}]$
we neglect the $\hat{k}$ term, since in $k=x_k p+\alpha_k Q' +k_t$ the
first piece gives zero contribution, and the second and the third one destroy
the logarithmic divergence of the $k_t$-integral.
So, indeed, the longitudinal polarization is entirely odd signature, and 
$g_1$ has only odd signature.

With this knowledge we return to the transverse target polarization
and show how the varios pieces resulting from (31) contribute to
the tensors in (2).
The first term containing $\hat{Q}$ leads to the spin structure: 
$Tr[\gamma_5 \hat{s} \gamma_{\nu} \hat{Q} \gamma_{\mu}]=4i
\epsilon_{\mu \nu \alpha \beta} Q^{\alpha} s^{\beta}$ which contributes to 
both $g_1$ and $g_2$.
 As said before, it
has odd signature. The second (even signature) term 
has the spin structure of $T_2$ ($g_2$):
$Tr[\gamma_5 \hat{s} \gamma_{\nu} \hat{k} \gamma_{\mu}]=4i
\epsilon_{\mu \nu \alpha \beta} k^{\alpha} s^{\beta}$, and in the DLA
we approximate $k_{\mu} \approx x_k p_{\mu}$ (cf. the discussion after (4)).
Finally the spin structure of the last (odd signature) piece,
$2(sk) Tr[ \gamma_5 \hat{k} \gamma_{\nu} \hat{Q}\gamma_{\mu}]$. 
In order to have
two logarithms we need a logarithmic integration $k_t^2 dk_t^2/(k_t^2)^2$.
For the transverse polarization $s_{\mu}=s^{\perp}_{\mu}$ we have already
a $k_t$ factor from the prefactor $(sk)$, and in the trace we therefore
have to retain the transverse component of $\hat{k}$. Recall that this piece
contributes to $g_{\perp} =g_1 +g_2$. On the other hand,
for the longitudinal polarization $s_{\mu}=p_{\mu}$
we put $(sk)=(pk)=\alpha_k s/2 =k_t^2/2$, 
and inside the trace we substitute $\hat{k}=x_k\hat{p}$. This contribution
is therefore smaller by the factor $x_k \approx x_{bj}$ than the `normal`
double logarithmic result for the longitudinal polarization. As a result,
this last (odd signature) piece is negligable inside $g_1$
and contributes only to $g_2$.
Combining all these results we arrive at the conclusion that $g_1$ has only
odd signature contributions whereas $g_2$ contains both even and odd signature
pieces. 
\begin{figure}
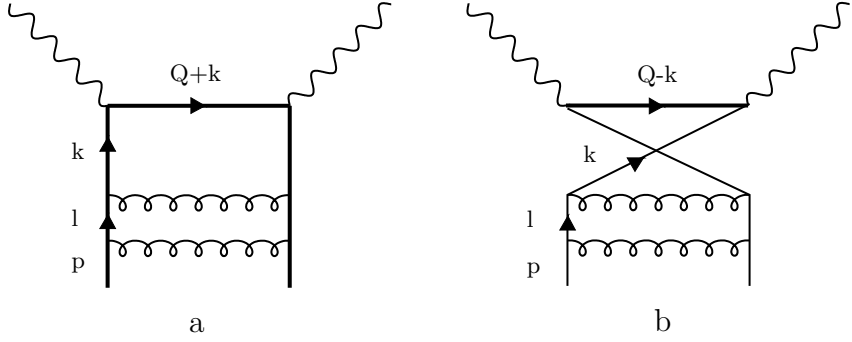

\begin{center}
\input paper56.fig8.pstex_t
\caption{Signature structure of the second loop}
\end{center}
\end{figure}

To complete our discussion we have to consider diagrams with one more 
rung (Fig.8). From the density matrix at the bottom we now have not only
the `old` matrix $\gamma_5 \hat{s}$ but also a term of the form
$(sl)\gamma_5 \hat{l}$. For the former we can repeat all previous steps.
For the latter, however, it is nontrivial to show that, at the double log 
level,
the term $(sl) Tr[\hat{k}\gamma_5\hat{l} \hat{k} \gamma_{\nu} \hat{k}
\gamma_{\mu}]$ does not 
contribute to the even signature case (otherwise not only $g_2$, but also 
$g_1$ would receive an even signature contribution). The even signature trace 
term is written in the form
\beqn
(sl) Tr[\hat{k}\gamma_5\hat{l} \hat{k} \gamma_{\nu} \hat{k}
\gamma_{\mu}] = k^2 (sl) Tr[\gamma_5 \hat{l} \gamma_{\nu}\hat{k} \gamma_{\mu}].
\eeqn
One possibility for obtaining the integral $dl_t^2/l_t^2$ is to keep
the transverse components in both $l$-factors. This gives:
\beqn
l_t^2 \frac{k^2}{2} Tr[\gamma_5 \hat{s} \gamma_{\nu} \hat{k} \gamma_{\mu}]
\approx l_t^2 \frac{k^2}{2} Tr[\gamma_5 \hat{s} \gamma_{\nu} x_k \hat{p} 
\gamma_{\mu}].
\eeqn
The other possibility is to retain, in the trace, the large ($x_l \gg x_k$)
longitudinal component of $l_{\mu}$. Instead of the rhs of (34) we obtain:
\beqn
k^2 (sl) Tr[\gamma_5 x_l \hat{p} \gamma_{\nu} \hat{k_t} \gamma_{\mu}].
\eeqn
At first sight this expression does not seem to have a double logarithm from
the $l_t$ integral. On the other hand, it is larger than (34) by the
ratio $x_l/x_k$, and we have to remember that there are corrections from
the propagators $1/k^2$. Putting $k^2=\alpha_k x_k s + k_t^2$ with
$\alpha_k=\frac{(k-l)_t^2}{x_ls}$ we expand:
\beqn
\frac{1}{k^2}& =& \frac{1}{k_t^2(1+\frac{x_k}{x_l} \frac{(k-l)_t^2}{k_t^2})}
        \nonumber \\
&\approx& \frac{1}{k_t^2} \left(1+\frac{2(k_t l_t)}{k_t^2}\frac{x_k}{x_l} +...
               \right).
\eeqn
Combining the correction term on the rhs with (35) we obtain the double 
logarithmic contribution:
\beqn
k^2 (sl_t) \frac{2(k_t l_t)}{k_t^2} Tr[\gamma_5 x\hat{p} \gamma_{\nu}
        \hat{k} \gamma_{\mu}] = \frac{k^2 l_t^2}{2} Tr[\gamma_5 x_k 
        \hat{p} \gamma_{\nu} \hat{s} \gamma_{\mu}], 
\eeqn
which exactly cancels the contribution from (34). This shows that, in fact,
also at the two loop level the structure $(sl)\gamma_5 \hat{l}$ leads to odd 
signature only.

From this discussion the following general pattern emerges. As we have shown 
before, in a general ladder diagram of Fig.6 each rung reproduces
the tensor structure $\gamma_5 \hat{s}$ and creates, in addition, the 
beginning a new branch. Follwing the first branch, we have a mixture of
even and odd signature, whereas the new branch belongs to odd signature
only. As a general result, during the evolution along any of these branches
signature is conserved. For the first branch, the final trace at the top 
of the ladder separates even and odd signature parts. In order to 
include the nonladder graphs in the correct way we have to discuss
even and odd signature branches separately.

Let us first follow the odd 
signature part of the branch ($\gamma_5 \hat{s}^{\perp}$): at each step it 
creates a new 
branch of pure odd signature with the spin structure $\gamma_5 \hat{k}$.
For any odd signature branch in Fig.7 (say, with the structure 
$\gamma_5 \hat{s}^{\perp}$ up to the j-th rung, and with the
structure $\gamma_5 \hat{k}$ above the j-th rung) we have the well-known
infrared evolution equation which takes into account both the ladder
graphs and the double logarithmic non-ladder bremsstrahlung contributions.
For each such branch we have the same result as for $g_1$:
\beqn
g_1(x,Q^2)=\frac{e_q^2}{2} 
          \int_{-i \infty}^{+i\infty} \frac{d\omega}{2\pi i} 
            \left(\frac{1}{x}\right)^{\omega}
            \left(\frac{Q^2}{\mu^2}\right)^{f^{(-)}(\omega)/2}
            \frac{\omega}{\omega-f^{(-)}(\omega)/2}.
\eeqn
Here $f^{(-)}(\omega)$ denotes the quark scattering amplitude: 
\beqn
f^{(-)}(\omega) = \omega -\sqrt{ \omega^2 -\frac{2C_F}{\pi} \alpha_s^L
            +\frac{2C_F}{2\pi^2 \omega}\alpha_s^N f^{(+)}_8}.
\eeqn
The non-ladder diagrams are contained in the function
\beqn
f^{(+)}_8=4\pi N \alpha_s^N \frac{\partial}{\partial \omega}
    \ln    \left( e^{z^2/4} D_{p}(z) \right),
\eeqn
where 
\beqn
z=\omega/\sqrt{\frac{N\alpha_s^N}{2\pi}} 
\eeqn
(for further details see the Appendix). 
Once we reach the top of the ladder, we `close` with the trace. The odd
signature part is induced by the $\hat{Q}$ part.

The correct counting of these odd signature branches is done in the 
following way. New branches are created only at ladder rungs. Non-ladder
gluons never change the spin structure. We therefore have to count the 
number of rungs, and we do this by formally distinguishing between
$\alpha_s^L$ (the strong coupling multiplying a gluon rung) and
$\alpha_s^N$ (the strong coupling multiplying a nonladder gluon).
This procedure turns the $\alpha_s$ in the second term in the square root 
expression (39) into $\alpha_s^L$, whereas the $\alpha_s$ in the third 
term of this square root, as well as in (40) and (41) becomes
$\alpha_s^N$.
As in (27), the number of new odd signature branches is obtained
by applying the derivative operator $-\alpha_s^L \frac{\partial}{\partial 
\alpha_s^L}$ onto $g_1$. In addition we have a contribution $g_1$ from the main
$\gamma_5 \hat{s}$ branch. So altogether we find:
\beqn
g_{\perp}^{odd}= \left(
1-\alpha_s^L \frac{\partial}{\partial \alpha_s^L} \right)
g_1|_{\alpha_s^L=
\alpha_s^N=\alpha_s}.
\eeqn
This generalizes the odd signature part of (28), in that in includes the
nonladder diagrams.

Next the even signature branch: it is contained only in the
tensor structure $\gamma_5 \hat{s}^{\perp}$, and it is projected out
by the $\hat{k}$ part of the final trace. Its coefficient 
is by a factor 2 smaller than in the usual $g_1$ calculation (since for the
longitudinal photon one can replace $\hat{Q}$ by $-2x\hat{p}$). For this
branch nonladder graphs cancel, so our result equals:
\beqn
- \frac{1}{2} g^L.
\eeqn
(cf.(25)).
 
Finally, in the one loop contribution we have to add also the nonladder
graph. In section 3 we have argued that in the diagrams of Fig.3d the two 
cuts across $q_1$ and $q_2$ cancel against each other. However, in the 
first loop (Fig.2b,c), where the cut 
of the quark $q_1$ line corresponds to $x=1$ and where we have the
term $\hat p\gamma_5\hat s$ in the initial quark density matrix,
the usual cancellation between the real and virtual DL
contributions does not work. Hence  
within the DLA we have to take care about the non-ladder
diagrams Fig.2b,c. They give a contribution of the form:
\beqn
\frac{3 e_q^2 C_F}{8\pi} \alpha_s.
\eeqn

Now we can add all contributions, (42), (43), and (44).
Since this result for the transverse polarization is proportional to
the sum $g_1+g_2$, we have to subtract the known result for $g_1$.
We obtain:
\beqn
g_2(x,Q^2)= -\frac{1}{2} g^L (x,Q^2) - \left(
\alpha_s^L \frac{\partial}{\partial \alpha_s^L} g_1 \right)|_{\alpha_s^L=
\alpha_s^N=\alpha_s} + \frac{3 e_q^2 C_F}{8\pi} \alpha_s.
\eeqn
Eq.(45) is our final result for $g_2$. It differs from the one 
obtained in ~\cite{ET} in several respects:
in ~\cite{ET} crossing properties and non-ladder graphs had not been analysed.
As a result, the even signature contribution $g^L$ was omitted, and
the second term did not correctly contain the nonladder graphs. Finally,
the last term representing the one loop contribution was missing.

\section{The Wandzura-Wilczek Relation}

Let us finally comment on the Wandzura-Wilczek (WW) relation (1) and see 
how our results fit into this equation. 
Since the WW-relation was proven to be valid only
for the twist-2 part of $g_2$ but not for the whole structure
function $g_2$, we do not expect that our DLA result for the small-$x$ 
behavior satisfies this identity. It is instructive to illustrate the
role of the twist-3 part in lowest order perturbation theory.
For small $\alpha_s\ll1$ both functions $g_1$ and $g_2$ behave as
const$+O(\alpha_s\ln Q^2)+O(\alpha_s\ln\frac1x)$, but for $g_2$
the ''const'' vanishes, and the term $O(\alpha_s$) goes to zero 
proportional to 
$x$. Therefore, at the order $\alpha_s$, the integral
on the r.h.s. of eq.(1) contains one extra $\ln1/x$, which on the lhs
can be contained only in the twist-2 part of $g_2$. But since 
the full $g_2$ (eq.(19)) has no such a logarithm at this order of
the coupling $\alpha_s$,
this logarithm in the twist-2 part must be cancelled by a similar logarithm in
the twist-3 part of $g_2$. This illustrates that in the small-x region the 
twist-2 and twist-3 contributions of $g_2$ cannot be separated in an easy way.

The systematic construction of the twist-2 and twist-3 operators
has, for example, been described in \cite{BKL}. The $n$-moment of 
the spin dependent structure function is given by the expectation value of the
operator:
 \begin{equation}
R=\left(\frac{2i}{Q^2}\right)^n Q'_{\mu_1}...Q'_{\mu_n}
\left[\bar{\psi}\gamma_5\gamma_
\sigma D_{\mu_1}...D_{\mu_n}\psi\right]\ ,
\end{equation}
where the matrix $\gamma_\sigma$ 
 is "directed" in accordance with the quark spin
 vector $s_\sigma$.
In order to separate the twist-2 and twist-3 components, we have to 
symmetrize. The most symmetric operator
\begin{equation}
R_1=\left(\frac{2i}{Q^2}\right)^nQ'_{\mu_1}...Q'_{\mu_n}\;\mbox{\bf S}
_{\{\sigma ,\mu_1,...,\mu_n\}}\left[\bar{\psi}\gamma_5\gamma_
\sigma D_{\mu_1}...D_{\mu_n}\psi\right]
\end{equation}
gives the leading, twist-2 contribution, while the
\begin{equation}
R_2=\left(\frac{2i}{Q^2}\right)^nQ'_{\mu_1}...Q'_{\mu_n}\;\mbox{\bf A}_
{[\sigma ,\mu_1]}\mbox{\bf S}_{\{\mu_1,...,\mu_n\}}
\left[\bar\psi\gamma_5\gamma_\sigma
D_{\mu_1}...D_{\mu_n}\psi\right]
 \end{equation}
corresponds to twist-3 (in eqs. (47) and (48) the symbols $\mbox{\bf
 S}_{\{\mu_1,..., \mu_n\}}$ and $\mbox{\bf A}_{[\sigma ,\mu_1]}$
denote symmetrization and antisymmetrization with respect
to the indices $\{\mu_1,...,\mu_n\}$ and $[\sigma
 ,\mu_1]$, respectively.) In the case of longitudinal
polarization of the target quark the vector $s_\sigma$ has the
same longitudinal direction as the derivative $D_\mu$.
Therefore the operator $R_2$ does not contribute to the longitudinal 
polarization, and twist-3 does not appear. This situation changes  
when one deals with the transverse spin vector $s^\perp_\sigma$. Now the
twist-2 contribution is given by the operator $R_1$, e.g. for $n=1$:
\begin{equation}
\Big<h|R_1|h\Big>\ \sim
Q'_\mu\Big<h|\bar\psi\gamma_5\gamma^\perp_\sigma (xp_\mu)\psi|h\Big>\ +
\ Q'_\mu \Big<h|\bar{\psi}\gamma_5\gamma_\mu k^\perp_\sigma\psi|h\Big>\ ,
\end{equation}
and $R_2$ becomes nonzero:
\begin{equation}
\Big<h|R_2|h\Big>\ \sim
Q'_\mu\Big<h|\bar\psi\gamma_5\gamma^\perp_\sigma (xp_\mu)\psi|h\Big>\ -
\ Q'_\mu \Big<h|\bar{\psi}\gamma_5\gamma_\mu k^\perp_\sigma\psi|h\Big>\ .
\end{equation}
Now it is easy to see that the Wandzura-Wilczek relation is valid only 
for the twist-2 part of the structure function. It follows just from the 
relativistic invariance \cite{BKL}; both sides are the matrix elements of the 
different components of the same twist-2 tensor operator.

The results of our calculation show that at small $x$ the WW-relation 
looks rather artificial. In particular, the twist-3 contribution is by no 
means small. For the first-loop we have shown that 
the twist-3 contribution adds an extra term to
$g_2$, which is even more singular at $x\to 0$. For the case of 
the one loop $O(\alpha_s)$ contribution the 
situation (and the behaviour of the moments of twist-2 and twist-3 
operators) has been discussed in detail in \cite{HZ}.
In the context of the MIT-bag model, a large twist-3 contribution to 
$g_2(x)$ has been emphasized in
~\cite{JJ}. As a result, we conclude that the WW relation cannot be used
to estimate the full function $g_2$ at small $x$.

\section{Conclusion}

In this paper we have calculated the small $x$ asymptotics of the spin
dependent structure function $g_2$ in the double log approximation of 
perturbative QCD. We started with the one and two loop approximations, and
we then succeeded to perform the sum over all orders in $\alpha_s$. 
The asymptotic small $x$ behaviour of
$g_2$ is governed by the rightmost singularity in
the $\omega$-plane at $\omega=\omega_0\simeq \sqrt{2C_F\alpha_s}$
$(\simeq0.4$ at $\alpha_s=0.2)$. This is the square root
singularity of $qq$-amplitude $f_0(\omega)$.
As a result, at $x\to0$ the structure function $g_2$ behaves as
\begin{equation}
g_2\ \propto\ x^{-\omega_0}\left(\frac{Q^2}{\mu^2}
\right)^{\omega_0/2}
\end{equation}
with the anomalous dimension $\gamma =\omega_0/2$ .

There is an interesting difference between our transverse
distributions $g_\perp(x,Q^2)$ (or $g_2(x,Q^2)$) and the
structure function $h_1(x,Q^2)$ which describes the evolution of
the operator $\langle\gamma_5\hat Q'\hat s^\perp\rangle$. 
The small $x$ behaviour of $h_1(x,Q^2)$ was studied in \cite{Ki}.
In the case of $g_2(x,Q^2)$ the double logarithmic ladder evolution
has been found to take place near the singularity at $j=0$, and
$g_2(x,Q^2)\sim x^{-\lambda}$,
with $\lambda=\omega_0\sim O(\sqrt{\alpha_s})>0$. In $h_1$, on the other hand,
the situation is quite different. Namely due to
the fact that after the interaction with the $s$-channel gluon the spin
structure vanishes 
($d_{\alpha\beta}\gamma_\alpha\gamma_5\hat Q'\hat
s^\perp\gamma_\beta = -4(\hat{Q}s^\perp)\gamma_5 = 0$),
the leading $\sim(1/x)^0$ part of the function $h_1$ has no DL evolution
at all, i.e. the singularity at $j=0$ in
the complex angular momentum plane does not contribute. Only the
singularity at $j=-1$ leads to the DL contribution, and the small-$x$
behavior of $h_1$ is suppressed by one power of $x$ compared to
$g_2$: $h_1(x,Q^2)=xf(\alpha_s\ln^21/x$,
$\alpha_s\ln1/x\ln Q^2)$.

Let us recall that in this paper, as a first step, we have calculated 
$g_2$ only for deep inelastic scattering on a quark target, i.e. 
only the evolution of the $g_2$ structure function within the 
perturbative QCD. Nothing has been said about
the initial distribution of $g_2$ coming from the confinement region.
On the other hand, in ref.\cite{RR} it has been argued that, within the 
parton model, the intrinsic transverse momenta of
the quarks (inside a proton) should lead to a non-zero
value of $g_2$ which satisfies the Wandzura-Wilczek relation
(1). However, in order to reproduce the relation (1) one has to use 
very large transverse momenta ($k_t^2\sim Q^2$; as the integral
over $k_t^2$ corresponds in \cite{RR} to the integration over
$dy$ in eq.(1) with $dy=dk_t^2/2m\nu$). Within the parton model such 
large values of intrinsic momenta $k_t$ look somewhat unnatural.

\section{Appendix}
\setcounter{equation}{0}

Here we will review the structure of the odd signature quark-quark amplitude 
and show where the QCD  coupling ($\alpha_s=g^2/4\pi$) 
corresponds to the ladder, and where it belongs to the non-ladder 
contributions. We have
to separate these contribution in order to take the derivative with respect to 
the ladder coupling $\alpha_s^L$ in the final expresion (45).

The well known equation for the odd signature quark-quark amplitude in the 
$\omega$ representation has the form\cite{KL}
\begin{equation}
f_0^{(-)} (\omega) = \frac{N^2-1}{2N} \frac{g^2_L}{\omega}
+ \frac{1}{8 \pi^2 \omega} \left( f_0^{(-)} (\omega) \right)^2
 -\frac{N^2-1}{N} \frac{g^2_{NL}}{4 \pi^2}
             \frac{1}{\omega^2} f_8^{(+)} (\omega)
\end{equation}
where the first two terms in the r.h.s. generate the ladder contribution 
while the last one 
 corresponds to the non-ladder graphs.

This last term on the rhs is due to the signature changing
contributions of gluon bremsstrahlung which lead us to define
an even-signature quark quark scattering amplitude $f_8^{(+)}(\omega)$
with color octet quantum number in the t-channel. This amplitude has
the infrared evolution equation 
\begin{equation}
f_8^{(+)} (\omega) = -\frac{g^2_L} { 2N \omega}
     +\frac{1} {8 \pi^2 \omega} \left( f_8^{(+)} (\omega) \right)^2
     +  \frac{N g^2_{NL}}{8 \pi^2 \omega} \frac{d}{d\omega} f_8^{(+)} (\omega)
\end{equation}
Its solution follows from the discussion given in ~\cite{KL}: using the
transformation
\begin{equation}
f_8^{(+)} (\omega) = Ng^2_{NL} \frac{\partial}{\partial \omega}
           \ln u(\omega)
\end{equation}
the Riccati equation (3) is equivalent to the linear differential
equation
\begin{equation}
\frac{d^2u}{dz^2} - z \frac{du}{dz} - 
\frac{1}{2N^2}\frac{g^2_L}{g^2_{NL}} u = 0
\end{equation}
where
\begin{equation}
z=\omega/\omega_0, \,\,\, \omega_0 = \sqrt{N g^2_{NL}/8 \pi^2}.
\end{equation}

After
a trivial transformation this equation is solved by a parabolic
cylinder function. As a result, $f_8^{(+)}$ has the form:
\begin{equation}
f_8^{(+)}(\omega)
          =  Ng^2_{NL} \frac{d}{d\omega} \ln \left( e^{z^2/4} D_p (z)
                              \right)
       \nonumber \\
         =  Ng^2_{NL} \frac{d}{d\omega} \ln
    \left(   \int_0^{\infty} dt t^{-1-p} e^{-zt-t^2/2}  \right)
\end{equation}
with
\begin{equation}
p=-\frac{1}{2N^2}\frac{g^2_L}{g^2_{NL}}.
\end{equation}
With this we return to (1) and obtain for $f_0^{(-)}$:
\begin{equation}
f_0^{(-)} = 4 \pi^2 \omega \left( 1 - \sqrt{ 1 - \frac{(N^2 - 1)}
            {4 \pi^2 N \omega^2} [g^2_L - \frac{g^2_{NL}}{2 \pi^2 \omega}
                                     f_8^{(+)} (\omega)]
                                                       } \right)
\end{equation}
(the minus sign in front of the square
root follows from the requirement that, for large $\omega$, the solution
has to match the Born approximation). \\ \\

\vspace{1cm}

{\bf Acknowledgement:} One of us (M.G.R.) gratefully acknowledges the
hospitality of the DESY Theory Group.

\end{document}